# Bibliometrics effects of a new paper level classification system based on reference reclassified by the origin of the papers citing them

**Marcos Peña-Rocha[1], Rocío Gómez-Crisóstomo[1], Vicente P. Guerrero-Bote[1*], Félix de Moya-Anegón[2]**

[1]Departamento de Información y Comunicación, Universidad de Extremadura, Badajoz, Spain

[2]SCImago Research Group, Granada, Spain

**\* Correspondence:**
Corresponding Author: guerrero@unex.es

**Abstract**

This study presents a comparative analysis between two scientific document classification systems. The first system employs the Scopus journal-based assignment method, adapted to a fractional model, while the second system uses an item-by-item system based on reclassified references according to the origin of the citers. The study's results are divided into three different sections: the first involves comparisons at the Scopus area level, the second examines comparisons at the category level, and the third tests various bibliometric indicators to identify the variations between the two systems. Highlighting the characteristics of the paper level system, it offers a reduction in the number of categories to which each document is assigned, achieving higher values of single-category assignment compared to the All Science Journal Classification (ASJC). When reclassifying areas and categories, the paper level system tends to accentuate differences at the extreme values, increasing the size of the largest categories and reducing that of the smallest ones. Moreover, the paper-by-paper system provides more homogeneous distributions in normalised impacts and adjusts values related to excellence more uniformly.

## 1  Introduction

A widely accepted classification by the scientific community is essential both to determine and quantify the progress made in each discipline and to normalize the different citation habits that exist among them.

Currently, the classifications used by bibliographic databases such as Web of Science (WoS) or Scopus are the most widely used for these purposes. These classifications operate at journal-level, where each paper is assigned to the category to which the journal in which it is published is ascribed.

Classifying documents based solely on the category assigned to a journal has some drawbacks that generate inaccuracies in bibliometric studies. One such drawback is associated with the multiple category assignments that journals have. As Álvarez-Llorente et al. (2023) state, "*not all the work a journal publishes are from all the categories to which it is assigned, indeed, quite the contrary is the*

*case*", causing inaccuracies since works are assigned to all the categories of the journal even though most belong to only one of them. This imprecision can even extend to both the classification of works and the normalisation of the scientific indicators previously mentioned.

Related to the previous issue, another inconvenience arises due to multidisciplinary categories. Journals classified under these categories publish works from various disciplines, but not all of their works are multidisciplinary. This situation creates a particular need to determine the specific disciplines to which each work belongs.

The initial attempts to address this aforementioned issue (Glänzel, Schubert, & Czerwon, 1999; Glänzel, et al., 1999) focused on the reference lists provided by the documents to classify them, indicating that references are the best way to define the content of a work. In 2021, Glänzel et al. (2021) proposed an improvement over the previous method by using a parameterised model that works with multiple generations of references.

In 2003, Glänzel and Schubert attempted to develop a classification system adapted for bibliometric use. In their study (Glänzel & Schubert, 2003), they proposed a hierarchical two-level classification system, developed through a three-step iterative process.

Years later, the implementation of clustering algorithms led to new methods for creating scientific classification schemes. By the late 2000s, several studies emerged based on hybrid clustering algorithms.

In (2008), Janssens et al. sought to classify the field of information science into different clusters. In a subsequent study, they used an analysis that combined citations with the full text of the article to classify WoS journals for the period 2002-2006 (Janssens et al., 2009). They concluded that using the combination of both components yielded more favourable results than treating them separately.

Boyack and Klavans (2010) tested the accuracy of several clustering algorithms on a large set of biomedical documents. The selected algorithms used co-citation analysis, bibliographic coupling, direct citations, and the combination of citations with the document's text to generate clusters. The last approach, which used a hybrid method based on bibliographic coupling, stood out by offering better results than the others.

The following year, another comparative study was conducted (Boyack et al., 2011) with a sample of over two million MEDLINE documents. This study compared the accuracy of nine clustering algorithms based on text similarity between documents with the information extracted from their bibliographic records. Using document titles and abstracts on one hand and MeSH thesaurus terms on the other, they developed a summary table that grouped the results generated by the different algorithms. They state that the indicator known as PMRA (PubMed Related Articles), described in 2007 (Lin & Wilbur, 2007) generates the most coherent clusters with a low computational cost. This indicator is used by the PubMed database to recommend articles related to the document selected by the users. Its functionality is based on the Poisson probabilistic distribution, using both MeSH thesaurus terms and the words from the abstracts and titles, with title words given twice the weight of those in the abstracts.

In 2012, researchers Waltman and van Eck (2012) conducted a study in which they presented a methodology for classifying approximately 10 million documents without requiring high computational cost, using direct citations as a method to determine the relationship between works. They concluded that using other types of bibliographic relationships, such as bibliographic coupling, could offer significant improvements. Additionally, they identified several issues to be addressed in future research, such as improving the labelling of areas, allowing publications to be assigned to more than one area, and incorporating new articles into an existing classification.

In another attempt to improve cluster precision, Boyack, Small, and Klavans (2013) sought to enhance the accuracy of co-citation clustering models by using a proximity analysis of references from a total of 270,521 full-text documents from 2007. After comparing the traditional co-citation model with the hybrid method, they concluded that the hybrid model achieves greater precision and reduces the size of the generated clusters. This finding is consistent with their previous studies, where hybrid algorithms produced more accurate results.

Although clustering algorithms provide solutions for classification needs, using the classifications they generate has significant drawbacks compared to established classifications. These algorithms create structures based on the papers available at the time of execution. Therefore, when incorporating new papers, it becomes necessary to run the algorithm again, which can result in a completely different structure and very different groupings. Additionally, there is the issue of labelling clusters, which is far from trivial.

For these reasons, none of the previous clustering proposals have been widely accepted by the scientific community. Consequently, in recent years, there have been several attempts to classify documents using schemes that are accepted by the scientific community, such as those outlined below.

Milojević (2020) aimed to reclassify WoS using the classification scheme employed by the database. Milojević's reclassification work involved a total of forty-five million documents, consisting of conference proceedings and scientific articles, each with at least one reference. The proposed method seeks to infer the category of an article through its references, with the goal of assigning documents to a single category. The first iteration of the system starts by using only references assigned to journals with a single category that is not multidisciplinary. Subsequent iterations include documents classified in previous passes. Milojević concludes that his method is easily replicable, does not require high computational costs, and avoids the problem of labelling categories by using a pre-established classification.

More recently, Álvarez-Llorente et al. (2024a) analysed a fractional classification system that uses the ASJC classification scheme from Scopus. This study is based on the multiple generations of references system proposed by Glänzel et al. (2021), but with the possibility of assigning up to five categories instead of three and using different assignment thresholds. This fractional classification system assigns weights to references based on the categories to which the journal containing the reference is assigned in Scopus ("weighted-counting") as opposed to the full-counting system in which all assignments have the same weight. In this study, they compare the results of the three schemes from the model proposed by Glänzel et al. (2021) in their work on multiple generations of references, where they are weighted according to the generation they belong to (denoted as M1, M2, and M3, depending on whether the references are first-generation, second-generation, or both, respectively).

Building on these schemes, new ones are generated where different parameters vary, such as the use of full or fractional counting of weights, the application of averaging or non-averaging counts to these systems, and the use of three different reference thresholds. For validation, they compare the results with the Scopus fractional assignment scheme (ASJC), with the multidisciplinary area and reallocated miscellaneous categories, as well as with the test collection known as "AAC" (Author's Assignation Collection) (Álvarez-Llorente et al., 2023), where the corresponding authors decide which categories best suit their works. They conclude by stating that higher thresholds for multiple category assignments, along with a system of fractional weights, yield better results. These parameters, using an averaged count with two generations of references, produce what they consider the most appropriate classification for bibliometric use (M3-AWC-0.8).

Even more recently, Álvarez-Llorente et al. (2024b) proposed a classification based on references reclassified according to the origin of the citers. This methodology reclassifies Scopus publications, but instead of using the category to which the journals, where each article's references are published, are assigned, it uses the category to which the journals of the citing articles belong. A total of twelve variants of the developed system are generated by modifying two parameters: the use of fractional or non-fractional weights and three different thresholds for multiple category assignments. By limiting the study to publications from the year 2020 in Scopus, it is possible to compare these variations with the ASJC system, the "M3-AWC-0.8" system, and the "AAC" test collection. After examining the results, the variant they found most suitable uses fractional weights with the highest threshold for multiple category assignments (U1-F-0.8), outperforming both ASJC and M3-AWC-0.8. They consider this variant's ability to assign a reduced number of categories per document using a fractional weight system, without the need for high computational requirements, to be a positive attribute.

With all this in mind, the aim of this study is to determine, through a comparative exercise over a broad period, the variation introduced by the paper level classification system developed by Álvarez-Llorente et al. (2024b) compared to the ASJC-based classification system from the Scopus database. This involves testing various indicators such as normalised impact, excellence distribution, and, in general, analysing how the data varies by examining the areas and categories that constitute the chosen classification. The following research questions are posed:

a) How does the distribution of documents across areas and categories change with the new classification?
b) How do documents flow between disciplines with a change in the classification system?
c) Does the fractional assignment of documents to categories differ between the two systems?
d) How does normalised impact vary depending on the system used?
e) Do the excellent documents differ in the new classification system?

## 2  Materials and methods

The data used in this study come from the Scopus database, specifically from a copy made in March 2024 to which SCImago has access through an agreement with Elsevier. This database employs a classification system known as the All Science Journal Classification (ASJC) (Elsevier, 2023).

This classification uses a two-level journal-based system, categorising them into 27 thematic areas, which are further subdivided into 311 categories. Among these areas, there is one designated as "*Multidisciplinary*," which is dedicated to purely multidisciplinary journals and does not have any subcategories.

The remaining 26 areas have a specific category dedicated to miscellaneous documents within them, comprising the name of the area followed by the word "*miscellaneous*" in parentheses. For example, within the "*Social Sciences*" area, there is a category named "*Social Sciences (miscellaneous)*".

Álvarez-Llorente, et al. (2024a) argued that reclassifying works with fewer than three references should be avoided as it would not provide a minimum level of significance. Figure 1 shows the annual percentage of papers registered in Scopus with fewer than three references.

This figure gradually decreases from 17.23% in 2003 to its minimum of 3.25% in 2022. During the period from 2003 to 2008, the number of documents with fewer than three references <u>exceeds</u> 10%.

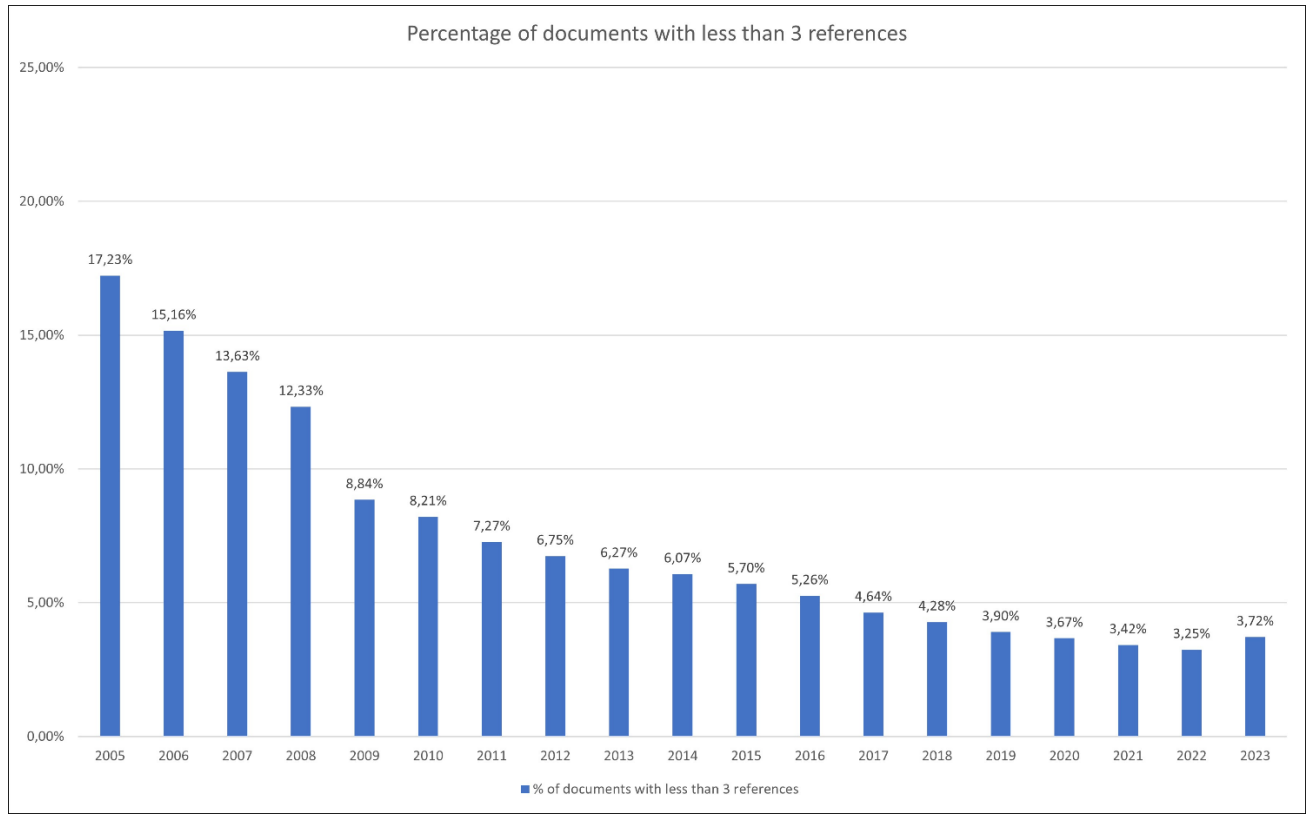

**Figure 1 Percentage of documents with less than 3 references**

A clear difference is observed between the early years of the period and the later years, with a notable drop of nearly 8.38% in the number of documents between 2005 and 2009.

Due to this, we decided to consider the last 12 years for the study (period 2012-2023), where the values are more stable and there are no significant peaks of difference between them. As a result, the total number of documents we finally worked with is 35,183,738 items.

To conduct a coherent comparison, we need to transform the ASJC classification into a fractional classification. For this purpose, we decided to follow the same method presented by Álvarez-Llorente et al. (2023) in which a fractional classification is derived from the ASJC of Scopus.

The scheme used eliminates the general "*Multidisciplinary*" area along with all miscellaneous categories from each area, resulting in a total of 26 thematic areas and 285 categories. However, to retain the values lost by removing the thematic area and miscellaneous categories, a division is performed where "*the weight assigned to an affiliation to the Multidisciplinary subject area is divided among the 285, and the weight assigned to the miscellaneous categories is divided among the rest of the categories of the same subject area*"(Álvarez-Llorente et al., 2023).

With respect to fractional assignment, the objective is to weight the number of assignments a journal has. This method ensures that the value of each assignment varies depending on the number of categories to which the journal is assigned; this value is the weight each assignment receives. The greater the number of assignments, the lower the weight of each individual assignment. Therefore, to determine the size of a specific class, we sum the values of all the weights that comprise it. We will refer to this sum of weights generally as documents.

The fractional classification system based on the ASJC scheme is compared throughout the study with a classification system based on reference reclassification, also of a fractional nature, with the peculiarity that the latter reclassifies references according to the origin of the citers. This system assigns up to a maximum of five categories to each document using a restrictive assignment threshold, thus avoiding excessive multiple category assignments. It is defined in detail in the work by Álvarez-Llorente et al. (2024b), and its implementation does not require high computational costs.

Below are some of the bibliometric indicators used in the study:

- Normalised Impact (NI): the average citation normalised received by each document, understood as the ratio between the citation received by the document and the average citation of documents of the same type, year, and category (Rehn & Kronman, 2008)
- Excellence at 10% (%Exc10): defined as the percentage of documents that belongs to the top 10% most cited of their same type, year, and thematic area (Bornmann et al., 2011).
- Excellence at 1% (%Exc1): defined as the percentage of documents that belongs to the top 1% most cited of their same type, year, and thematic area (Bornmann et al., 2011).

## 3 Results

To present the results in a structured way, we begin by examining the data related to the 26 knowledge areas used by Scopus. Then, we move on to the second part, which focuses on the 285 categories into which the data is organised. Finally, we review the analysis of several bibliometric indicators such as normalised impact and the different types of excellence.

### 3.1 Areas

Figure 2 compares the documents accumulated by each of the classification systems analysed in this study. The presented graph consists of the 26 knowledge areas used by the Scopus database, each of which is represented by two columns, with each column further divided into two parts.

The green area included in both columns represents the common weight shared by both systems, while the left column, coloured in blue, represents the part that is unique to the ASJC system, and the right column, coloured in orange, shows the part that is unique to the paper-by-paper system.

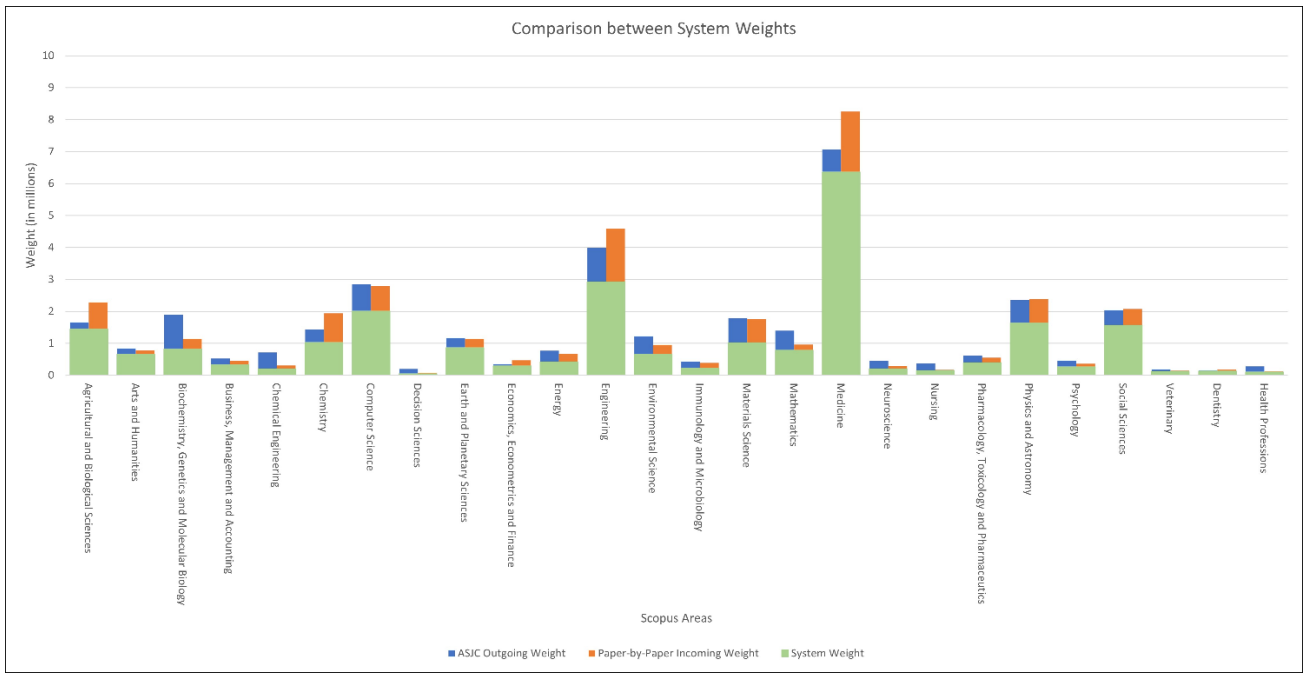

**Figure 2 Comparison between the weights of each Scopus area according to the system used**

Illustrated below is a distribution of the flow network between Scopus areas, showing the flow of documents that occurs when transitioning from the ASJC system to the new classification, developed using the "*SCImago Graphica*" tool (Hassan-Montero et al., 2022).

This tool allows the representation of networks as shown in Figure 3, where the "*LinLog*" algorithm (Noack, 2007, 2009) is applied to position the nodes, taking into account the arcs and their weights. Both the nodes and the arcs are scaled according to their weights.

The weight of the nodes is determined by the number of documents assigned to them by the paper-by-paper classification system. Meanwhile, the weight of the links represents the documents that transition from one system to another, specifically from the ASJC classification to the paper level classification.

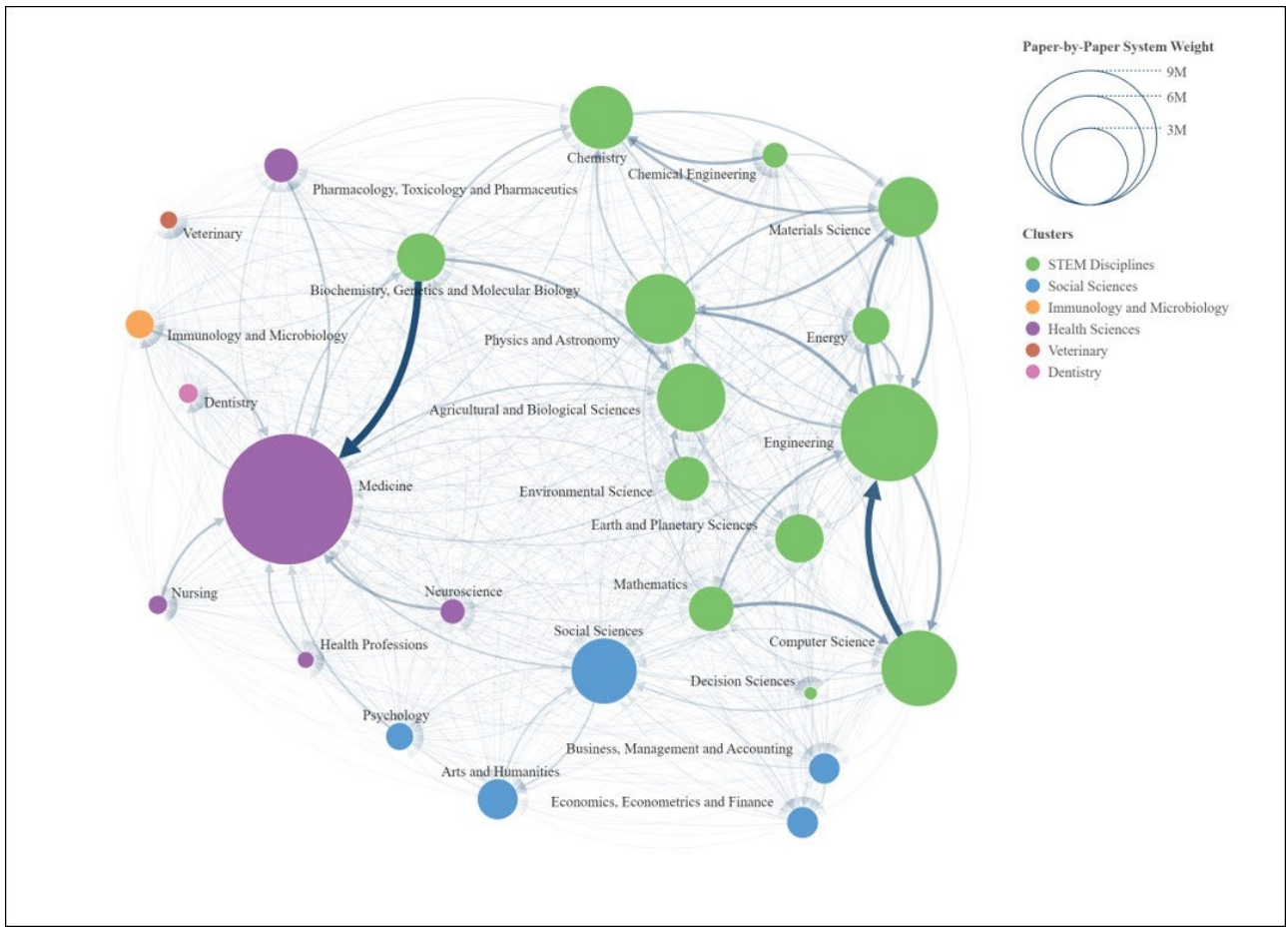

**Figure 3 Scopus Areas Network Diagram**

The clusters identified by the program are generated using the community detection algorithm by Clauset (Clauset et al., 2004), which distinguishes six main clusters:

- STEM Disciplines (Science, Technology, Engineering and Mathematics)
- Social Sciences
- Immunology and Microbiology
- Health Sciences
- Veterinary
- Dentistry

The first cluster is composed of thirteen areas, with the largest being "*Engineering"*, which has a weight of 4,590,331 documents, followed by "*Computer Science*", with a total of 2,790,049 documents. The strongest flow of information within a single area occurs in this cluster, from "*Computer Science*" to "*Engineering*".

The cluster dedicated to the Social Sciences is composed of four main areas, with the largest being "*Social Sciences*", which has a weight of 2,070,013 documents.

Next is the cluster dedicated to Health Sciences, composed of six nodes, with the largest being "*Medicine*", which has 8,247,437 documents, making it the largest node in the network. Within this grouping of areas, the most notable link is from the STEM Disciplines cluster, specifically from the area of "*Biochemistry, Genetics and Molecular Biology*" to "*Medicine*".

Finally, the clusters dedicated to Veterinary, Immunology, and Dentistry consist of a single area, named "*Veterinary*", "*Immunology and Microbiology*" and "*Dentistry*", respectively.

To highlight the most significant links between the areas, Table 1 lists the links with a weight exceeding 100,000 documents. This table comprises three columns: the first indicates the node from which the link originates, the second indicates the node to which the link leads, and the third shows the strength of the link.

The areas of *"Medicine"*, "*Engineering"*, and "*Materials Science*" receive the most weight, with the area of "*Medicine*" receiving the heaviest link from the area of "*Biochemistry, Genetics and Molecular Biology*", with a weight of 554,677 documents.

For the case of second strongest link, it also has a considerable weight transfer, being the one from "*Computer Science*" to "*Engineering*" with 487,984 documents. The link in the opposite direction ranks as the sixth strongest, with a value of 237,567 documents.

Continuing with the study of the thematic areas, a scatter plot is included, showing the Scopus areas according to the percentage of incoming or outgoing documents they have with respect to the paper-by-paper system (Figure 4). In this representation, the nodes with a higher percentage of incoming weight are positioned to the right, while the areas with a higher percentage of outgoing weight are placed at the top.

The concept of "*outgoing documents*" refers to documents that were previously classified under an area in the ASJC system but are no longer included in that area after changing systems. Conversely, "*incoming documents*" represent the number of documents that, under the new classification system, are assigned to a different area of knowledge than they were in the ASJC classification, thus becoming part of the new area.

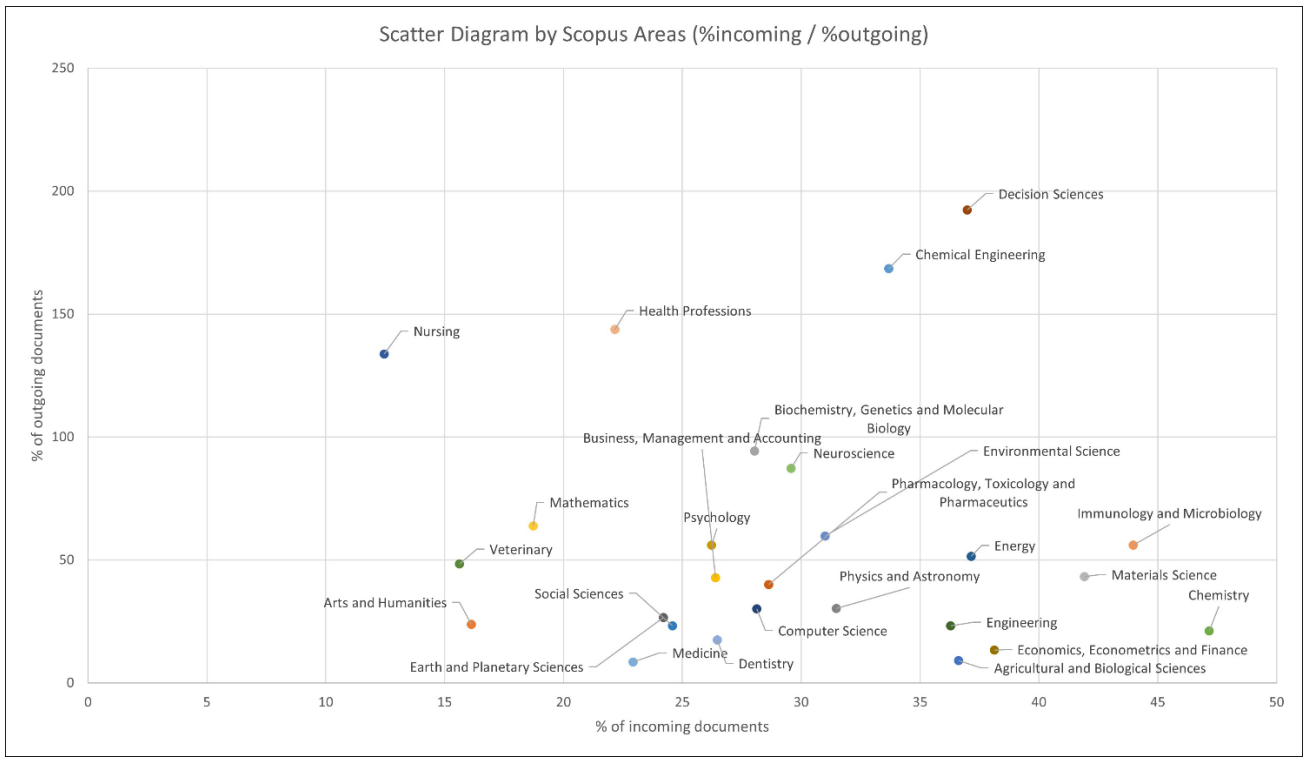

**Figure 4 Scatter Diagram by Scopus Areas with the percentage of incoming and outgoing documents**

We highlight a section of the graph where more than half of the areas are grouped; this occurs between the ranges of 20% to 40% incoming weight and 0% to 60% outgoing weight, encompassing a total of 14 areas.

To conclude with the data related to the areas, the average weight of each Scopus area according to the classification system used is analysed, along with the percentage of documents that belong exclusively to that category. Figure 5 uses two different axes of representation:

- The left axis ranges from 0% to 100% and is represented by blue and orange columns. The blue column indicates the percentage of documents assigned to a single category in the ASJC system, while the orange column represents this percentage in the paper level system.
- The right axis ranges from 0 to 1 and is represented in the figure by blue circles and orange triangles. This axis represents the average weight of each document within the area.

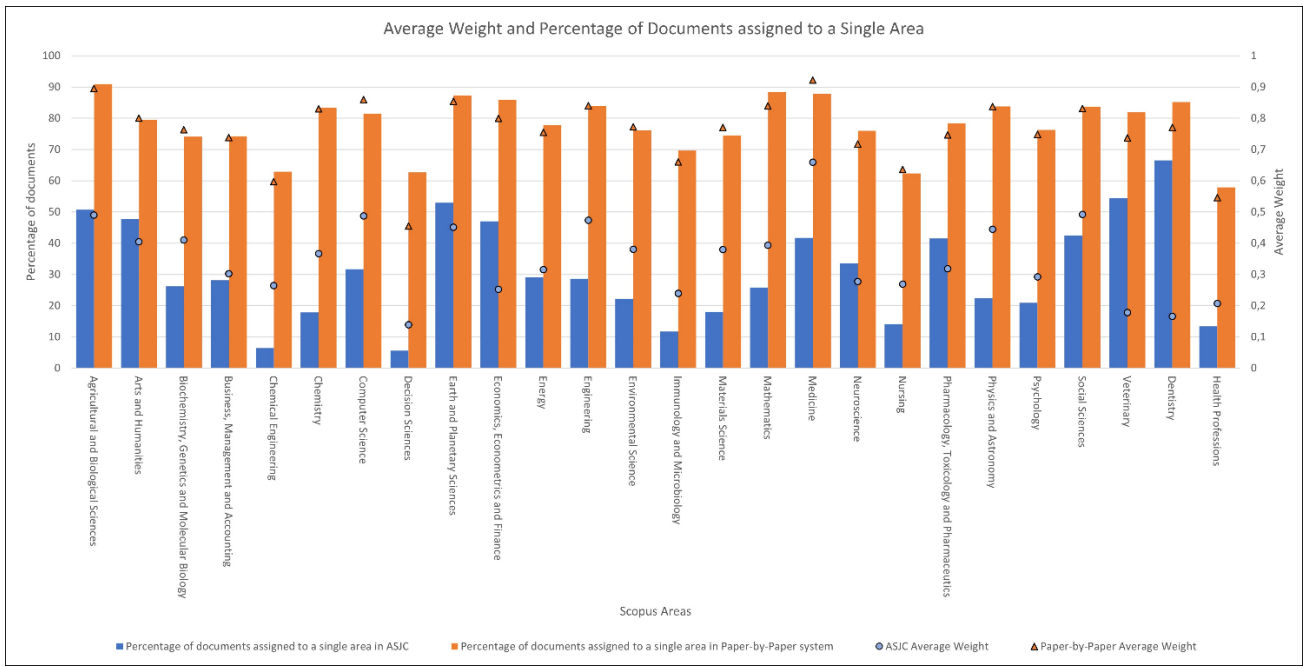

**Figure 5 Average weight and percentage of documents assigned to a single area according to the system used**

Starting with the columns, it is observed that the percentage values of assignment to an area in the ASJC classification are lower than those in the new classification. These values range from 5.61% to 66.49% for the ASJC classification, whereas for paper level system they range from 57.88% to 90.90%.

In the case of average weights per document, it is also notable that the values are higher with the new system. For the ASJC classification system, the area of "*Medicine*" stands out with an associated average weight of 0.65, followed by the area of *"Social Sciences"* with a value of 0.49 and the areas of "*Agricultural and Biological Sciences*" and "*Computer Science*" with a value of 0.48.

Similar to the journal-based classification, the paper-by-paper system also shows that the area with the highest average weight per document is "*Medicine*" with a value of 0.92, followed by the areas of "*Agricultural and Biological Sciences*" and "*Computer Science*" with values of 0.89 and 0.85, respectively.

Due to the weights of each document being less fractionated in the paper level classification compared to the ASJC classification, the percentage of documents assigned to a single category is higher in the new system. It can be observed that the average weights in the new classification system are consistently higher than those in the old system.

### 3.2   Categories

Next, a bar chart is included to visually represent the distribution of categories by their weight according to the system used. Figure 6 shows this diagram, which is divided into intervals of 100,000 documents. The blue columns represent the ASJC classification, while the orange bars represent the paper level system.

This diagram, functioning as a histogram, illustrates how many of the Scopus categories fall within each interval according to the number of documents they contain.

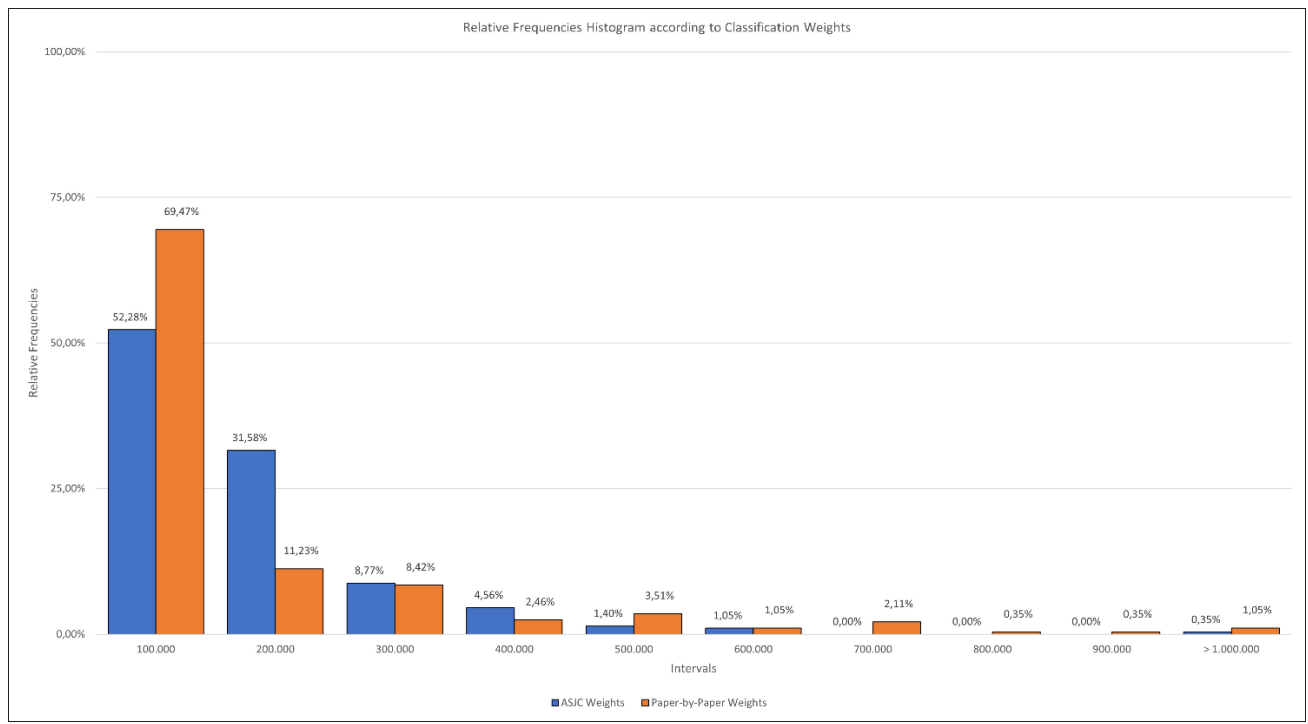

**Figure 6 Relative frequencies histogram according to classification weights**

The two systems share similarities, highlighting how in both cases more than 50% of the categories are grouped within the first interval. For the ASJC classification, 52.28% of the categories consist of fewer than 100,000 documents, with the next bracket containing 31.58% of the categories in the 100,000 to 200,000 document range. With the new classification system, nearly 70% of the categories are composed of fewer than 100,000 documents, with a total of 69.47%.

The new classification system has a peculiarity when dealing with categories that are at the extremes of size: it tends to decrease the smaller categories and further increase the larger ones. This is why the percentage of categories with fewer than 100,000 documents are almost 17% higher in the new classification system compared to the ASJC.

In both cases, the representations have a negative trend, starting with the highest point in the first interval and progressively decreasing. There are some intervals that do not contain any categories, these are the ones that have a value of zero.

It is important to highlight that there is a specific category in the paper-by-paper classification system that stands out due to its highest value. The category in question is "*Electrical and Electronic Engineering*," which has a total weight of 2,027,561 documents. The next category is "*Oncology*", with a total of 1,022,225 documents.

To show the general trend of the categories concerning information flows, the results are grouped in a table (Table 2). The first part is dedicated to the weights of the categories according to the classification system used, analysing the average and coefficient of variation statistics. The second part deals with the outgoing and incoming flows with respect to the paper level classification.

Starting with the differences between the classification systems, the first notable point is that, although the average number of documents in both systems is the same, the coefficient of variation is not. Documents in the ASJC classification system are distributed among the categories more homogeneously, with a variation of 93.31% compared to the 163.51% variation in the paper level system.

For the cases of outgoing and incoming flows, a similar pattern is observed. Despite the average of both variables being 67,424 documents, the coefficient of variation is higher for incoming flows than for outgoing flows. Given these high variation values, it can be stated that the new system increases the size of the larger categories and reduces the size of the smaller categories.

Both the average and the coefficient of variation for outgoing flow are higher than for incoming flow. The average value for incoming flow is 37.26%, while for outgoing flow it reaches 253.23%. Additionally, the incoming flow is more homogeneous, with a value of 52.11% compared to the 150.28% for outgoing flow. This is because, when working with percentage values, the difference is much more pronounced when there is an outgoing flow from smaller categories compared to when the incoming values are directed towards larger categories, resulting in a smaller percentage difference.

Similar to Table 1, another table (Table 3) is designed to show the significant flows between categories, including all those that exceed 40,000 documents.

The categories of "*Oncology*" and "*Electrical and Electronic Engineering*" receive the most weight, with the field of oncology occupying the first position and the next four positions held by the latter category. The heaviest link is from "*Cancer Research*" to "*Oncology*", with a weight of 135,709 documents. The areas associated with these fields of knowledge correspond to the areas linked by the heaviest connection in Table 1.

The top-ranked link has a noteworthy peculiarity: the incoming and outgoing categories belong to different communities according to the algorithm used in Table 3. The outgoing category is part of the "*Biochemistry, Genetics and Molecular Biology*" area, which is included in the STEM disciplines cluster, while the incoming category falls within the "*Medicine*" area, in the Health Sciences cluster.

Finally, Table 4 is included to explain the differences in category weights according to the classification system. It is divided into two parts:

- The first part is dedicated to the ASJC system, which includes the average weight of Scopus categories and their average percentage of assignment to a single category, along with their respective coefficients of variation.
- The second part is dedicated to the paper-by-paper system, where the structure is similar but this time with the data from the new system.

Focusing on the variables of the assigned weights, it is evident that the value of the categories is, on average, lower in the ASJC system, with an average of 0.059 documents per category compared to 0.324 documents in the paper-by-paper system. Despite both systems registering a very similar coefficient of variation for this variable, the weight assigned to each category is higher on average in the paper-by-paper system, resulting in less fractional distribution of weights among the different categories.

Regarding the variable that measures the percentage of documents assigned to a single category, the superiority of the paper-level system over the ASJC is notable, with more than half of the documents being assigned to a single category compared to 13.51% in the ASJC classification. Additionally, the coefficient of variation in the ASJC system is 96.74%, while in the paper-level system it is 44.69%, providing not only higher average values but also more homogeneous ones.

## 3.3 Bibliometrics indicators

To assess how the change in classification affects quality indicators, we compare the values provided by the normalised impact quality indicator and the excellence indicator.

Figure 7 represents the annual average evolution of the absolute value of the difference in each paper between the normalised impact calculated from the two classifications.

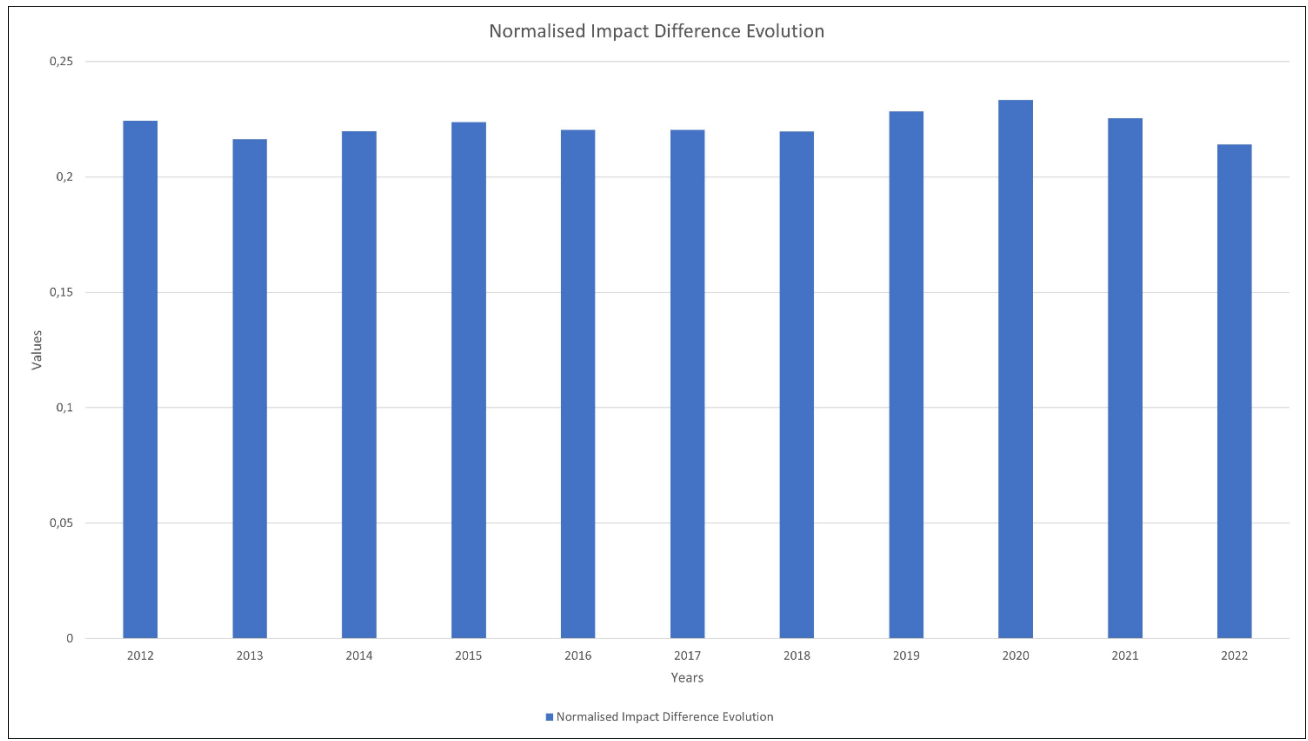

**Figure 7 Normalised Impact difference evolution**

Since citation data are not comparable across disciplines due to variations between them, it is important that the normalised impact indicator tries to homogenise the citation habits of similar documents.

The paper level system shows a 22% variation in the average normalised impact compared to the traditional ASJC system, indicating that the normalised impact varies significantly depending on the classification used.

Although the rest of the figures extend the study up to the year 2023, this year is excluded from this figure because it has had a shorter citation window, making the values less precise than those of the previous years.

Figure 8 shows a comparison by areas, represented by a column chart that illustrates the standard deviation of the normalised impact according to the classification used. The variable "*STD NI ASJC*" is represented in blue, indicating the standard deviation for the ASJC classification, while the variable "*STD NI Paper-by-Paper*" is represented in orange, indicating the standard deviation for the paper level classification.

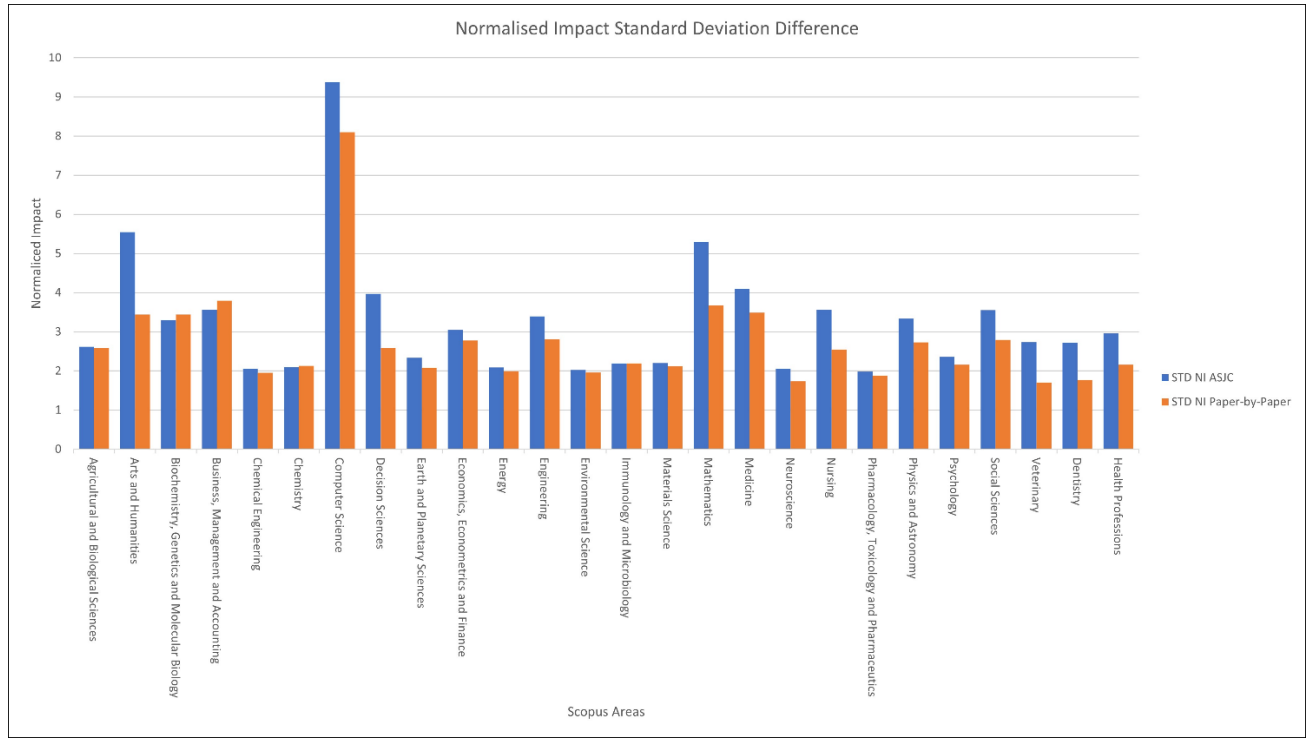

**Figure 8 Normalised Impact Standard Deviation Difference**

In the ASJC classification, high values are notable in areas such as "*Computer Science*", "*Arts and Humanities*", and "*Mathematics*". Meanwhile, in the paper-by-paper classification, the areas with the highest standard deviation in normalised impact are "*Computer Science*", "*Business, Management and Accounting*" and "*Mathematics.*"

The results show that the paper level classification tends to offer lower standard deviation values compared to the ASJC classification, indicating that it is more homogeneous in terms of normalised impact.

For the excellence indicators, we work with the indicators described in the methodology: *%Exc10* and its variants in Figure 9, and *%Exc1* and its variants in Figure 10.

Starting with Figure 9, consists of three columns: the first, in green, labelled as *%Exc10*, represents the documents that are excellent in both classifications systems. In orange, we define "*%Exc10 non-ASJC*" as the percentage of documents that were not considered excellent in the ASJC classification but are now considered excellent with the paper-by-paper classification. Conversely, "*%Exc10 ASJC*" is defined in blue as the percentage of documents that were considered excellent in the ASJC classification but are not considered excellent in the paper-by-paper classification.

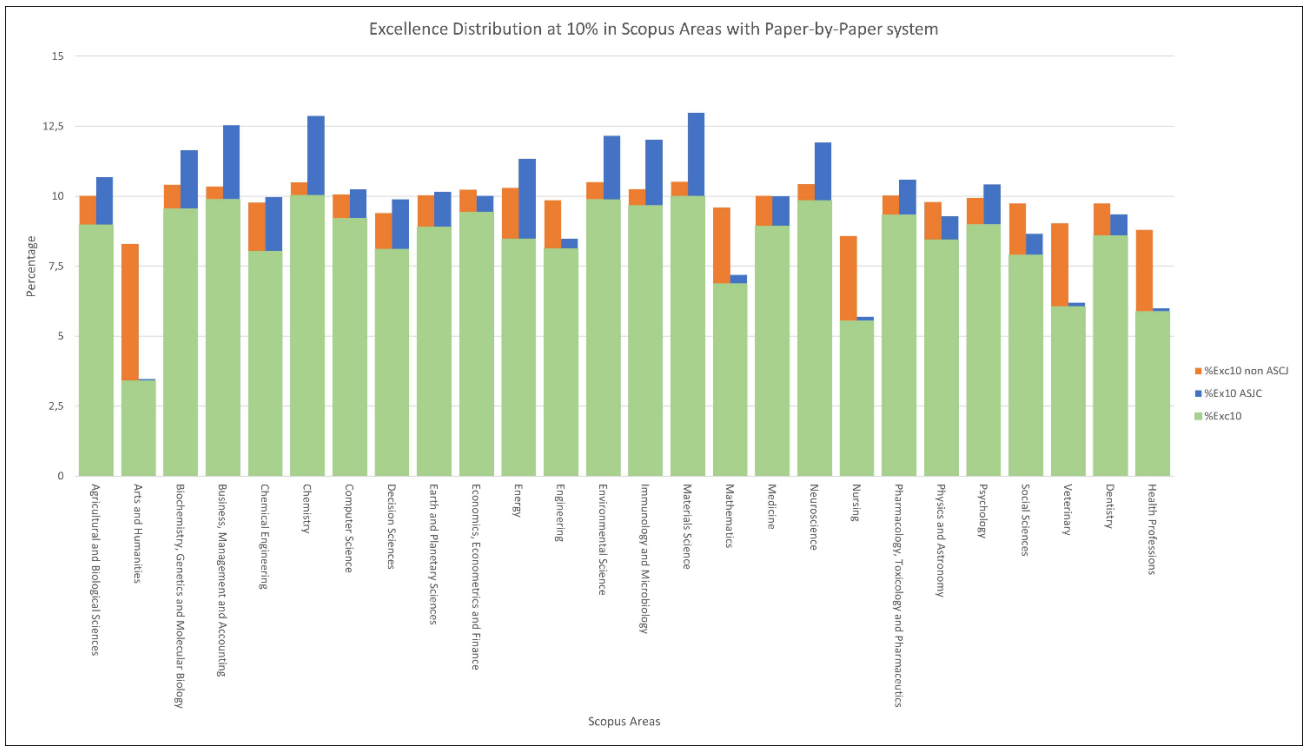

**Figure 9 Excellence distribution at 10% in Scopus areas with paper-by-paper system**

It is evident that the categories most benefited by the new classification are "*Arts and Humanities*", "*Nursing*", "*Veterinary*" and "*Health Professions*," showing the most notable differences between their "*Exc10 non-ASJC*" and "*Exc10 ASJC*" values.

There is a relationship between the areas with a smaller number of documents in the top 10% of excellence and those that achieve higher values in the paper level classification. If the areas are ordered from lowest to highest by their "*%Exc10*", we find that the same areas that benefited from the classification change, mentioned in the previous paragraph, are the ones with the lowest "*%Exc10*" values.

In other words, the areas with a lower percentage of documents belonging to the top 10% of their area correspond to the areas with the highest percentage in the indicator of documents that become excellent in paper level but were not considered excellent in ASJC.

For the case of 1% excellence, the process is similar to the previous one, but focusing on the top 1% of works, as shown in Figure 10. With this indicator, the four areas with the lowest "*%Exc1*" are again the four previously mentioned, which then benefit the most in "*%Exc1 non-ASJC*".

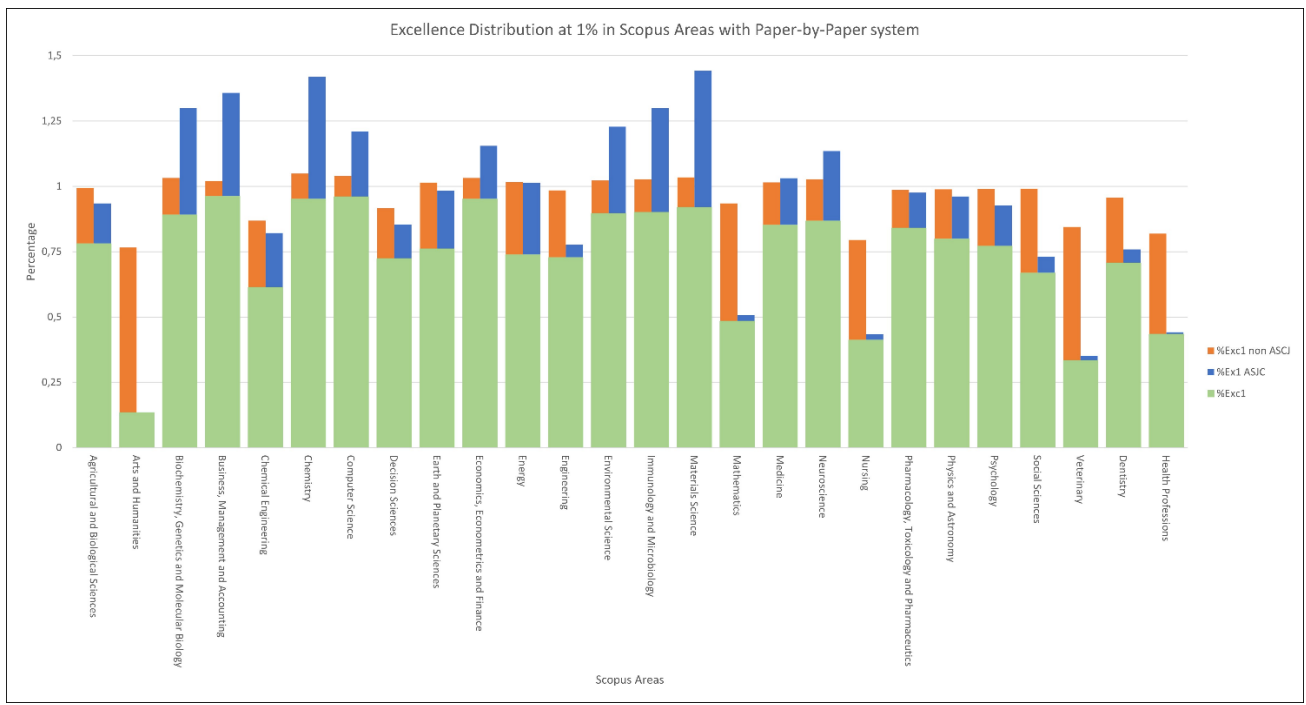

**Figure 10 Excellence distribution at 1% in Scopus areas with paper-by-paper system**

Although the "*Arts and Humanities*" category achieves very high excellence values in the new classification for both types of excellence, this is directly related to the way bibliometric indicators are measured in these disciplines. The dissemination of knowledge through articles is not as frequent as in purely scientific disciplines, so these data should be treated separately.

## 4      Conclusions

After conducting a comparative analysis between the two classification systems, several conclusions can be drawn based on the different aspects addressed. Starting with the distribution of documents across disciplines, the paper-by-paper system, far from balancing the size of the ASJC areas and categories, tends to accentuate differences at the extremes, increasing the size of both the largest areas and categories while decreasing the smallest ones.

Logically, the areas composed of a larger number of documents are the ones that incorporate (or lose) the most documents, as seen in the graphical representation in Figure 3, where the areas of "*Medicine*" and "*Engineering*" are the largest. In percentage terms, areas such as "*Decision Sciences*" or "*Health Professions*", which are the smallest in size, have very high percentages of outgoing documents. Similarly, the areas of "*Chemical Engineering*" and "*Nursing*" exhibit this trend but with lower values. These are two fields of knowledge that are among the smallest in terms of size, but the percentage of outgoing documents exceeds 100%.

Another area worth highlighting, in this case for the percentage of incoming documents, is "*Immunology and Microbiology*". Despite being the ninth smallest area, it achieves the second-highest percentage of incoming documents, with nearly 45%.

This characteristic of increasing size disparities is not desirable for scientometrics, but it may highlight the lesser significance of some categories, where very few works cite the intellectual foundations specific to them.

The new system, compared to ASJC, reduces the number of fractions into which each document is divided. This characteristic aligns with the findings of Álvarez-Llorente: "*the assignment of a reduced number of categories per paper is a highly desirable feature*" (Álvarez-Llorente et al., 2024b).

Since the weights are less fractionated in the new system, the values for assignment to a single area in the ASJC classification have an average of 30%, while with the paper level system, this value increases to nearly 78% of the documents. In all areas, a higher percentage of documents are assigned to a single category in the new system.

This is indeed a desirable characteristic because it better represents reality. As previously indicated, there are multidisciplinary journals that publish works from various disciplines, but far fewer works are genuinely multidisciplinary or interdisciplinary.

For the case of normalised impacts, there is a significant variation between them depending on the classification used. As shown in Figure 7, the value resulting from calculating the variable using the annual average evolution of the absolute value of the difference between the normalised impact of the paper-by-paper system and the ASJC system indicates that, on average, this figure varies by 22%.

This new classification offers a more homogeneous distribution of normalised impacts, grouping documents with similar citation habits together. The standard deviation values shown by the ASJC classification system average 3.24, compared to 2.71 in the new system. This characteristic is also desirable because one of the main uses of scientometric classifications is to compare works belonging to disciplines with the same citation habits, and this indicates that in the new system, the citation habits within each category are more homogeneous.

Regarding the issues related to excellence indicators, there is a common percentage of excellent documents that remain in both classifications: for documents belonging to the top 10% of excellence, this figure is 8.72%, while for the 1% excellence indicator, this value is 0.79% of documents in common.

Finally, we can state that the number of excellent documents in each area varies depending on the classification used. The new system ensures that the areas that do not reach the 10% or 1% excellence thresholds get closer to the respective values of 10 and 1. Logically, this characteristic of the new system is also desirable.

In conclusion, we believe that the new system, overall, represents in a better way the reality and exhibits more desirable characteristics.

**Funding**

This work has been funded by Ministerio de Ciencia e Innovación de España (MCIN), Agencia Estatal de Investigación (AEI) grant project PID2020-115798RB-I00. It has also been funded through the grants for pre-doctoral contracts for the formation of PhDs: PRE2021-098826, funded by MCIN/AEI/10.13039/501100011033 and FSE+.

**Conflict of interests**

The authors declare that the research was conducted in the absence of any commercial or financial relationships that could be construed as a potential conflict of interest.

**Tables**

**Table 1 Stronger Links between Scopus Areas**

| Outgoing Flow | Incoming Flow | Number of Documents |
|---|---|---|
| Biochemistry, Genetics and Molecular Biology | Medicine | 554,677 |
| Computer Science | Engineering | 487,984 |
| Engineering | Materials Science | 251,137 |
| Mathematics | Computer Science | 244,036 |
| Materials Science | Engineering | 237,686 |
| Engineering | Computer Science | 237,567 |
| Physics and Astronomy | Engineering | 237,242 |

| Chemical Engineering | Chemistry | 207,289 |
| Materials Science | Physics and Astronomy | 202,166 |
| Materials Science | Chemistry | 190,198 |
| Mathematics | Engineering | 186,563 |
| Biochemistry, Genetics and Molecular Biology | Agricultural and Biological Sciences | 182,003 |
| Neuroscience | Medicine | 176,740 |
| Environmental Science | Agricultural and Biological Sciences | 169,053 |
| Energy | Engineering | 156,043 |
| Nursing | Medicine | 150,441 |
| Engineering | Physics and Astronomy | 149,653 |
| Physics and Astronomy | Materials Science | 148,419 |
| Immunology and Microbiology | Medicine | 132,625 |
| Pharmacology, Toxicology and Pharmaceutics | Medicine | 124,767 |
| Health Professions | Medicine | 117,915 |
| Physics and Astronomy | Chemistry | 117,182 |
| Biochemistry, Genetics and Molecular Biology | Chemistry | 116,030 |
| Chemistry | Materials Science | 111,797 |
| Medicine | Biochemistry, Genetics and Molecular Biology | 102,320 |

**Table 2 Summary measures of incoming and outgoing flow**

*Summary Measures of Incoming and Outgoing Flow*

| | Variables | Average | Variation Coefficient |
|---|---|---|---|
| Classification System | Weight in the ASJC system | 123,451 | 93.31% |
| | Weight in the paper-by-paper system | 123,451 | 163,51% |
| Outgoing Flow | Outgoing Flow | 67,424 | 81.40% |
| | Percentage of outgoing flow with respect to paper-by-paper | 253.23% | 150.28% |
| Incoming Flow | Incoming Flow | 67,424 | 191.85% |
| | Percentage of incoming flow with respect to paper-by-paper | 37.26% | 52.11% |

**Table 3 Stronger Links between Scopus Categories**

| Outgoing Flow | Incoming Flow | Number of Documents |
|---|---|---|
| Cancer Research | Oncology | 135,709 |
| Energy Engineering and Power Technology | Electrical and Electronic Engineering | 79,875 |
| Computer Science Applications | Electrical and Electronic Engineering | 77,289 |
| Electronic, Optical and Magnetic Materials | Electrical and Electronic Engineering | 72,584 |
| Control and Systems Engineering | Electrical and Electronic Engineering | 69,991 |
| Computer Networks and Communications | Electrical and Electronic Engineering | 59,130 |
| Space and Planetary Science | Astronomy and Astrophysics | 59,085 |
| Electronic, Optical and Magnetic Materials | Condensed Matter Physics | 58,381 |
| Finance | Economics and Econometrics | 55,233 |
| Instrumentation | Electrical and Electronic Engineering | 53,710 |
| Mechanics of Materials | Mechanical Engineering | 53,101 |
| Instrumentation | Atomic and Molecular Physics, and Optics | 49,930 |
| Electronic, Optical and Magnetic Materials | Atomic and Molecular Physics, and Optics | 49,286 |
| Neurology | Neurology (clinical) | 49,241 |
| Condensed Matter Physics | Electrical and Electronic Engineering | 48,398 |
| Theoretical Computer Science | Software | 47,388 |
| Building and Construction | Civil and Structural Engineering | 46,408 |
| Hardware and Architecture | Computer Networks and Communications | 44,986 |
| Information Systems | Computer Networks and Communications | 44,957 |
| Atomic and Molecular Physics, and Optics | Condensed Matter Physics | 43,897 |
| Signal Processing | Electrical and Electronic Engineering | 41,395 |
| Physical and Theoretical Chemistry | Organic Chemistry | 40,959 |
| Computer Science Applications | Computer Networks and Communications | 40,602 |
| Hardware and Architecture | Electrical and Electronic Engineering | 40,378 |

**Table 4 Summary measures of categories weights**

*Summary measures of Categories Weights*

| Classification System | Variables | Average | Variation Coefficient |
|---|---|---|---|
| ASJC | Weight assigned to categories | 0.059 | 69.22% |
| | Percentage belonging to a single category | 13.51% | 96.74% |
| Paper-by-Paper system | Weight assigned to categories | 0.324 | 66.37% |
| | Percentage belonging to a single category | 56.08% | 44.69% |